\documentclass[%
 aip,
rsi,%
 amsmath,amssymb,
 reprint,%
]{revtex4-1}

\usepackage{graphicx}
\usepackage{dcolumn}
\usepackage{bm}

\begin{document}

\title{Optical pumping of quantum dot micropillar lasers}

\author{L. Andreoli}
\email{louis.andreoli@femto-st.fr}
\affiliation{D\'{e}partement d'Optique P. M. Duffieux, Institut FEMTO-ST,  Universit\'e Bourgogne-Franche-Comt\'e CNRS UMR 6174, Besan\c{c}on, France.}%

\author{X. Porte}
\affiliation{D\'{e}partement d'Optique P. M. Duffieux, Institut FEMTO-ST,  Universit\'e Bourgogne-Franche-Comt\'e CNRS UMR 6174, Besan\c{c}on, France.}%

\author{T. Heuser}
\affiliation{Institut für Festk\"{o}rperphysik, Technische Universit\"{a}t Berlin, Hardenbergstraße 36, 10623 Berlin, Germany.}%

\author{J. Grosse}
\affiliation{Institut für Festk\"{o}rperphysik, Technische Universit\"{a}t Berlin, Hardenbergstraße 36, 10623 Berlin, Germany.}%

\author{B. Moeglen-Paget}
\affiliation{D\'{e}partement d'Optique P. M. Duffieux, Institut FEMTO-ST,  Universit\'e Bourgogne-Franche-Comt\'e CNRS UMR 6174, Besan\c{c}on, France.}%

\author{L. Furfaro}
\affiliation{D\'{e}partement d'Optique P. M. Duffieux, Institut FEMTO-ST,  Universit\'e Bourgogne-Franche-Comt\'e CNRS UMR 6174, Besan\c{c}on, France.}%

\author{S. Reitzenstein}
\affiliation{Institut für Festk\"{o}rperphysik, Technische Universit\"{a}t Berlin, Hardenbergstraße 36, 10623 Berlin, Germany.}%

\author{D. Brunner}
\affiliation{D\'{e}partement d'Optique P. M. Duffieux, Institut FEMTO-ST,  Universit\'e Bourgogne-Franche-Comt\'e CNRS UMR 6174, Besan\c{c}on, France.}%

\date{\today}

\begin{abstract}

Arrays of quantum dot micropillar lasers are an attractive technology platform for various applications in the wider field of nanophotonics.
Of particular interest is the potential efficiency enhancement as consequence of cavity quantum electrodynamics effects which makes them prime candidates for next generation photonic neurons in neural network hardware.
However, in particular for optical pumping their power-conversion efficiency can be very low.
Here we perform an in-depth experimental analysis of quantum dot microlasers and investigate their input-output relationship over a wide range of optical pumping conditions.
We find that the current energy efficiency limitation is caused by disadvantageous optical pumping concepts and by a low exciton conversion efficiency.
Our results indicate that for non-resonant pumping into the GaAs matrix (wetting layer), 3.4\% (0.6\%) of the optical pump is converted into lasing-relevant excitons, and of those only 2\% (0.75\%) provide gain to the lasing transition.
Based on our findings we propose to improve the pumping efficiency by orders of magnitude by increasing the aluminium content of the AlGaAs/GaAs mirror pairs in the upper Bragg reflector.

\end{abstract}

\maketitle

\section{\label{sec:Introduction}Introduction}
In recent years arrays of vertically emitting lasers have experienced a rebirth based on many novel applications such as 3D sensors for mobile phones \cite{Seurin2016}, infrared illumination \cite{Zhou2014}, free-space communications \cite{Haghighi2020} and high-power lasers for material processing and optical pumping \cite{Seurin2010}.
Crucially, hardware-based artificial neural networks are an effervescent field that can benefit significantly from such arrays in the future \cite{Brunner2013a,Brunner2015,Heuser2020}. 

Photonic neural networks have recently received substantial attention \cite{VanDerSande2017}.
One of their essential ingredients are energy efficient optical nonlinear elements acting as neurons, as well as creating parallel photonic interconnects \cite{Moughames2020}.
In this context, arrays of quantum-dot micropillar lasers (QDMLs) are promising candidates for artificial neurons.
Among the key advantages of vertically emitting semiconductor lasers for neuromorphic computing \cite{Vatin2019,Robertson2020} are multi-GHz signal processing bandwidths \cite{Brunner2013a}, a high photonic neuron density enabled by their small size footprint and in principle low threshold pump powers \cite{Heuser2018}. 
The latter is a direct consequence of the high spontaneous emission coupling efficiency $\beta$ between the lasing mode and quantum dot (QD) gain inside high-quality microcavities \cite{Yamamoto1991,Gerard1998,Chow2018}. 

Vertically emitting lasers arrays are attractive for several applications, and electrically pumped QDMLs have achieved 20\% power conversion efficiency \cite{Reitzenstein2008}.
However, individually addressing electrically pumped QDMLs that are densely packed into a large array requires an elaborate electrode layout \cite{Heuser2020b}, and upscaling to arrays of hundreds of individually contacted QDML seems to be technologically out of reach.
Much of such a laser chip's surface-area would need to be assigned to electrical connections, which limits the lasers' rf-modulation bandwidth and prohibits dense integration.
3D fabrication could mitigate the challenge of dense integration, however, the increased capacity of signaling wires in such architectures results in even stronger bandwidth limitations.

In contrast, optical pumping of semiconductor microlasers provides several advantages and remains highly relevant in the context of, both fundamental research and emerging technological applications.
In fact, large numbers of densely packed lasers can be pumped using diffractive multiplexing in free-space \cite{Maktoobi2020}, and 3D printed waveguides \cite{Moughames2020} have the potential to integrate such a system.
In either setting, external modulation of the optical pump enables switching-speeds unattainable with electrical pumping.
In highly application relevant optically pumped vertical external cavity surface-emitting lasers demonstrate exceptional brightness \cite{guina2017optically}, and similar properties could be attained with coherently locker arrays of microlasers.

Dense arrays comprising hundreds of optically-driven QDMLs that emit in the near-infrared region ($\sim 980$~nm) have recently been demonstrated \cite{Heuser2018}.
For neural network applications reducing the array's inhomogeneous broadening is paramount \cite{Bueno2016}, and this important technological challenge has recently been successfully addressed \cite{Heuser2020}.
However, the power conversion efficiency of optical pumping is unfortunately less than 1\% \cite{Reitzenstein2006}, mainly because of absorption losses in the upper DBR.
This creates a serious obstacle for fundamental research and application relevance alike. 

Here, we experimentally study optical pumping of AlGaAs/GaAs micropillar lasers with InGaAs quantum dots in the active layer in detail and examine their power-conversion efficiency.
We operate eight QDMLs with two conceptually different pumping scenarios, in which we pump either wetting layer (WL) states at $\lambda^{\textrm{P}}\approx 915~$nm or transitions in the GaAs matrix at $\lambda^{\textrm{P}}\leq786~$nm, in both case at a range of wavelengths.
The pumping energies of both concepts are separated by more than 220~meV.
They therefore operate in different absorption and reflection ranges of the QDML's mirrors and their exciton-relaxation mechanisms traverse different quantum states.
By including these effects in a phenomenological model on the pumping efficiency we are able to determine the relative conversion ratios which ultimately determine the laser's global efficiency.
Importantly, our extended statistical study is based on over 150 input-output curves obtained under different conditions, which lends robustness to our results.
The lasers are distributed across an array of 20$\times$20 QDMLs \cite{Heuser2020}.

Studies on QDMLs have focused so far mainly on fundamental emission properties such as high-$\beta$ lasing \cite{Kreinberg2019} and single-QD lasing effects \cite{Gies2017}.
In these studies the power-conversion efficiency was of secondary importance and the layer and device design focused mainly on achieving pronounced light-matter interaction in the frame of cavity quantum electrodynamics \cite{Lermer2012}.
Two aspects of current optically pumped QDML devices design are mostly responsible for their poor power-conversion efficiency.
The first is that only a small fraction of the optical pump is actually absorbed at a position where it can contribute to population inversion.
This is due to (i) strong optical absorption of pump photons in the GaAs layers of the laser's top mirror when pumping with photon energies above the GaAs bandgap, and (ii) due to the small pump absorption coefficient of the WL.
The second issue is that the conversion of absorbed pump photons into QD-excitons partaking in lasing is very inefficient.
For pumping the GaAs intra-cavity layer we demonstrate that this efficiency is only $\sim$2$\%$, for pumping the WL this value further decreases to $\sim$0.7$\%$.
These numbers indicate that relaxation of high-energy bulk GaAs excitons happens via additional decay channels into the QDs besides relaxation via the WL.

Importantly, there is much room for improvement making only use of established techniques and tweaks to the usual sample design.
These have the potential to significantly enhance the relevant pump absorption as well as the excitation conversion efficiency.
Thus, by developing optimized layer designs the realization of optically pumped QDML with far higher pump-conversion efficiency should be possible.
Finally, we also provide robust measurements of the lasers' $\beta$ factor thanks to our statistics.
Here, we find that WL pumping is clearly beneficial compared to pumping bulk GaAs: $\beta^{\textrm{WL}}\approx0.1$ vs. $\beta^{\textrm{GaAs}}\approx0.02$.
Our results are of general relevance, as they quantify generic properties linked to DBRs within the GaAs material-system, as well as QD as gain material in micro-cavities.

\section{\label{sec:QDMPlasers}Quantum dot micropillar lasers}

Our quantum dot micropillar laser arrays (QDMLA) are fabricated in successive growth and etching steps.
A planar microcavity sample is epitaxially grown by means of metal-organic chemical vapor deposition.
The layer design of the planar microcavity consists of a central one-$\lambda$ thick GaAs cavity, sandwiched between a lower and an upper distributed Bragg reflector (DBR) respectively composed of 27 and 23 $\lambda/4$-thick Al$_{0.9}$Ga$_{0.1}$As~/~GaAs mirror pairs.
Three stacked layers of InGaAs QDs with an estimate density of $1\times10^{10}$~cm$^{-2}$ are centered inside the QDML's one-$\lambda$ thick laser cavity ($L^{\textrm{LC}}=279~$nm), and QD s-states provide optical gain by light-matter interaction with the fundamental cavity mode.
The growth of self assembled semiconductor QDs is associated with the formation of a thin quantum well, referred to as the wetting layer (WL). Subsequently, the vertical micropillars are nanoprocessed by high-resolution electron beam lithography in combination with reactive-ion etching~\cite{Heuser2020}.
This processing steps result in dense arrays of 20x20 QDMLs with a spacing of only 8.3~$\mu$m and a spectral homogeneity of $\Delta\lambda=$0.14~nm.
More information about this particular array can be found in~\cite{Heuser2020}.
We placed our sample in a liquid nitrogen cryostat with $\textrm{T}^\textrm{o}_{\textrm{sample}} = 130$~K measured below the sample. 

\begin{figure*}[t]
	\centering
	\includegraphics[width=11cm]{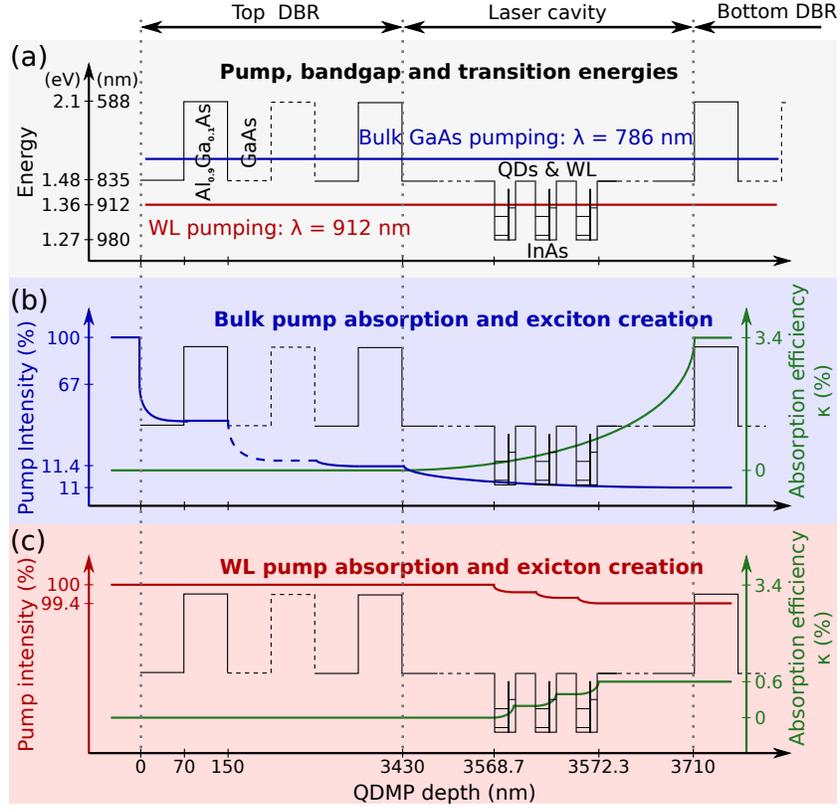}
	\caption{
		Schematic illustration of the relevant mechanisms for optically pumping QDMLs.
		Panel (a) gives the energies of wetting later and bulk GaAs pump lasers, bulk material bandgaps as well as WL and QD transitions. Panel (b) and (c) illustrate the pump intensity as pump photons traverse the sample, and the resulting relevant exciton creation efficiency for WL and bulk GaAs pumping, respectively. The left (right) y-axis in (b) and (c) give the local pump intensities (pump-photon to pump exciton conversion ratio).}
	\label{fig:EnergyDiagramQDMPscheme}
	\hrule
\end{figure*}

\section{\label{sec:OptPump}Optical pumping of QDML}

Optical pumping relies on a cascade of exciton creation, relaxation and their capture by the optical gain medium, here the QDs.
This multi-sequence process is therefore non-trivial, and determining the efficiency of each fundamental process with a potential link to material properties is essential for the understanding and further device optimization.
In this way the obtained insight becomes transferable between different laser structures and potential modifications to the material system.
The main processes governing how optical pumping with power $P^{\textrm{P}}$ finally drives the lasing transition are schematically illustrated in Fig. \ref{fig:EnergyDiagramQDMPscheme} and are discussed in the following.

\subsection{\label{sec:PumpAbs} Pump absorption}

\begin{figure}[t]
	\centering
	\includegraphics[width=8cm]{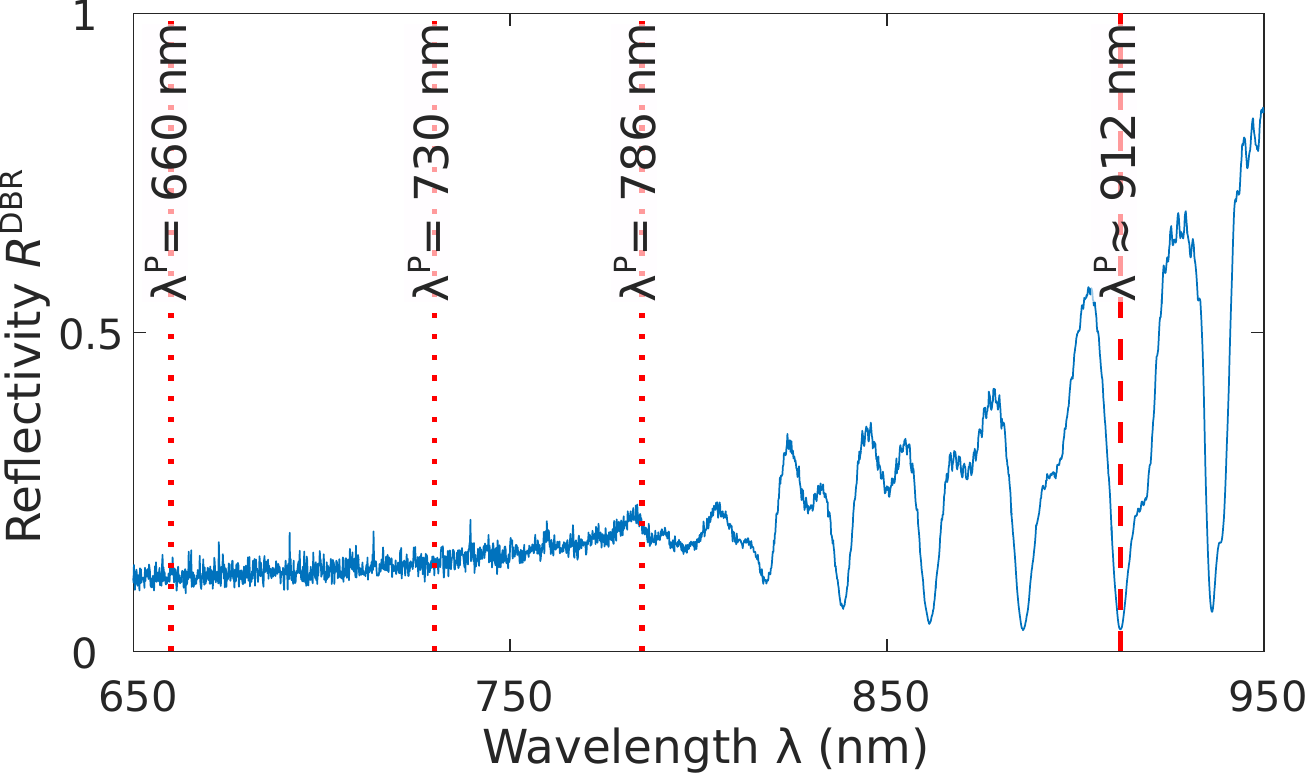}
	\caption{
		Reflection spectrum of the DBR. Pumping wavelengths for the wetting layers (WL) and GaAs are indicated by the red dashed or the dotted lines, respectively.
	}
	\label{fig:DBRSpectrum}
	\hrule
\end{figure}

The incident pump laser must pass the top DBR, whose transmission $T^{\textrm{DBR}}(\lambda)$ depends on wavelength sensitive reflection of the DBR stop-band $R^{\textrm{DBR}}(\lambda)$ and optical absorption (mainly in the GaAs layer) $A^{\textrm{DBR}}(\lambda)$.
While we do not have direct access to $T^{\textrm{DBR}}(\lambda)$, we can determine $R^{\mathrm{DBR}}(\lambda)$ via reflectivity measurements, see Fig. \ref{fig:DBRSpectrum}(a).
Furthermore, $A^{\textrm{DBR}}(\lambda)=1-e^{-\alpha^{\textrm{GaAs}}(\lambda,\textrm{T}^\textrm{o}_{\textrm{sample}}) L^{\textrm{DBR}}}$ can be precisely calculated using the accurately determined GaAs absorption coefficients $\alpha^{\textrm{GaAs}}(\lambda,\textrm{T}^\textrm{o}_{\textrm{sample}})$ and the combined thickness $L^{\textrm{DBR}}=1603~$nm of the DBR's GaAs layers.
Assuming that we can neglect other sources of mirror losses, we obtain 
\begin{equation}\label{eq:DBRtransmission}
	T^{\textrm{DBR}}(\lambda) = [1-R^{\textrm{DBR}}(\lambda)]e^{-\alpha^{\textrm{GaAs}}(\lambda,\textrm{T}^\textrm{o}_{\textrm{sample}}) L_{\textrm{DBR}}} .
\end{equation}

We investigate pumping the cavity's bulk GaAs at $\lambda^{\textrm{P}}=786~$nm, $\lambda^{\textrm{P}}=730~$nm or $\lambda^{\textrm{P}}=660~$nm.
Optically pumping in this range has an undesired side effect: the pump energy is above the laser-cavity's \emph{and} the DBR GaAs layer's band gap, see Fig. \ref{fig:EnergyDiagramQDMPscheme}(a).  The consequence is strong absorption by the DBR's GaAs layers, see Fig. \ref{fig:EnergyDiagramQDMPscheme}(b).
Quantum tunneling of excitons created inside the DBR towards the QDs is strongly suppressed by the energy-barrier of the DBR's Al$_{0.9}$Ga$_{0.1}$As layers, and these excitons are therefore not contributing to the lasing process.  The absorption of GaAs at 130~K for our three pump wavelengths are $\alpha^{\mathrm{DBR}}(786~\textrm{nm},130 ~\textrm{K})=1.248~\mu\textrm{m}^{-1}$, $\alpha^{\mathrm{DBR}}(730 ~\textrm{nm},130 ~\textrm{K})=1.532~\mu\textrm{m}^{-1}$ and $\alpha^{\mathrm{DBR}}(660 ~\textrm{nm},130 ~\textrm{K})=2.615~\mu\textrm{m}^{-1}$.
Here, we used the data of reference \cite{Aspnes1986}, which we extrapolated to 130~K according to reference \cite{Sturge1962}.  Furthermore, we cannot neglect the DBR's reflection at these three wavelengths, see dotted lines in Fig. \ref{fig:DBRSpectrum}. Details on how the DBR-reflection spectrum was obtained will be provided in Sec. \ref{sec:Experiment}.  Combining DBR reflection and absorption, we obtained $T^{\mathrm{DBR}}(786~\textrm{nm})=0.114$,  $T^{\mathrm{DBR}}(730~\textrm{nm})=0.074$ and $T^{\mathrm{DBR}}(660~\textrm{nm})=0.014$.

In contrast, WL states are below the band gap of the DBR's materials,  cf. Fig. \ref{fig:EnergyDiagramQDMPscheme}(a), which leads to $\alpha^{\textrm{GaAs}}(\lambda>830,130)\approx0$.
Conveniently, a DBR reflection minimum at $\lambda^{\textrm{P}}\approx912~$nm, cf. Fig. \ref{fig:DBRSpectrum}(a) dashed red line, is in the vicinity of the WL transition and we measured $R^{\textrm{DBR}}(912.3~\textrm{nm})=0.025$.
DBR mirror losses are therefore only determined by reflection, leading to $T^{\mathrm{DBR}}(912.3~\textrm{nm})=0.975$ at the WL's wavelength range.

Photons transmitted by the DBR arrive at the QDMP laser cavity where they need to be converted into upper level excitons through absorption according to $A^{\textrm{g}}(\lambda^{\textrm{P}})=1-e^{-\alpha^{\textrm{g}}(\lambda^{\textrm{P}},130) L^{\textrm{g}}}$.
Here, $\alpha^{\textrm{g}}(\lambda^{\textrm{P}},130)$ and $L^{\textrm{g}}$ are the absorption coefficient and length of the gain medium, respectively. The result is an absorbed pump power
\begin{equation}\label{eq:PumpAbs}
	P^{\textrm{P}}_{\textrm{abs}}(\lambda^P) = \kappa(\lambda^{\textrm{P}}) P^{\textrm{P}},
\end{equation}
\noindent with $\kappa(\lambda^{\textrm{P}})=T^{\textrm{DBR}}(\lambda^{\textrm{P}}) A^{\textrm{g}}(\lambda^{\textrm{P}})$ as the effective pump absorption.

Pumping the bulk GaAs of the QDML's one $\lambda$-thick lasing cavity corresponds to  $\alpha^{\textrm{g}}(\lambda^{\textrm{P}},130~\textrm{K}) = \alpha^{\mathrm{DBR}}(\lambda^{\textrm{P}},130~\textrm{K})$, and $L^{\textrm{g}} = L^{\textrm{LC}}=279~$nm.
This results in $\kappa^{\textrm{GaAs}}(786~\textrm{nm}) = 3.43\cdot10^{-2}$, $\kappa^{\textrm{GaAs}}(730~\textrm{nm}) = 2.64\cdot10^{-2}$ and $\kappa^{\textrm{GaAs}}(660~\textrm{nm}) = 0.71\cdot10^{-2}$.
In the case of WL pumping, absorption relevant for optical gain only happens in the three WLs with mono-layer thickness.
The InAs absorption coefficient is $\alpha^{\textrm{P}}(912~\textrm{nm},130~\textrm{K})=3.4~\mu\textrm{m}^{-1}$, and the three WLs below the QD layers have a combined thickness of three monolayers, i.e. $L^{\textrm{P}}=1.6~$nm.
The consequence of this very thin absorbing layer for gain is a low $\kappa(\approx912~\textrm{nm}) = 6\cdot10^{-3}$, see Fig. \ref{fig:EnergyDiagramQDMPscheme}(c).
Considering these numbers it is clear that the non-optimized QDML sample design strongly limits their power-conversion efficiency which can therefore not exceed 3$\%$ in the case of GaAs pumping due to the strong absorption losses in the top DBR.
Interestingly, while the top DBR is mostly loss-less for pumping the WL, the monolayer thin absorption section for gain limits the QDMP's efficiency to below 0.6$\%$.

\subsection{\label{sec:LaserRate} Laser rate equations}

To obtain a detailed understanding of the QDML emission properties we apply a laser-rate equation description\cite{Agrawal1993}, which yields the input-output of a QDML \cite{Reitzenstein2006,Yamamoto1991}
\begin{equation}
	P^{\textrm{P}}_{\textrm{abs}}(\lambda^P) = \frac{h\nu^ {\textrm{L}}\gamma}{\beta\delta} \biggl[ \frac{n^{\textrm{L}}}{1+n^{\textrm{L}}}(1+\xi)(1+\beta n^{\textrm{L}})-\xi \beta n^{\textrm{L}} \biggr],
	\label{eq:LIequation}
\end{equation}
\noindent with $\nu^{\textrm{L}}$ as the QDML's lasing frequency, the cavity decay rate $\gamma$ and $n^{\textrm{L}}$ as the photons in the laser mode of the QDML.
The spontaneous emission factor $\beta$ is the fraction of the global spontaneous emission that is collected by the lasing mode.
The efficiency at which upper level excitons are converted into QD s-state excitons providing gain is $\delta$, and parameter $\xi = n_0 \beta / \gamma \tau_{\textrm{sp}}$ additionally includes the number of excitons at transparency $n_0$ and the spontaneous exciton lifetime $\tau_{sp}$.

Equations~(\ref{eq:PumpAbs}) and (\ref{eq:LIequation}) link the number of lasing photons $n^{\textrm{L}}$ to the pump power $P^{\textrm{P}}$ via fundamental laser parameters $\gamma$, $\beta$ and $\delta$.
Finally, the QDML's output power $P^{\textrm{L}}$ is given by
\begin{equation}\label{eq:QDMLpower}
	P^{\textrm{L}} = v_{\textrm{g}} \eta^{\textrm{DBR}} h \nu^{\textrm{L}} n^{\textrm{L}},
\end{equation} 
with group velocity $ v_{\textrm{g}}$.
We calculate the DBR's outcoupling losses via \begin{equation}
	\eta^{\textrm{DBR}} = \frac{1}{L^{\textrm{LC}}} \textrm{log} \biggl( \frac{1}{\sqrt{ R^{\textrm{DBR}}(\lambda^{\textrm{L}}) }} \biggr) ,
\end{equation}
\noindent and the top DBR's reflectivity at the lasing wavelength $R^{\textrm{DBR}}(\lambda^{\textrm{L}})=0.9957$ is calculated based on the number of DBR mirror pairs and their refractive index.

\subsection{\label{sec:FreePara} Free parameter determination}

The QDML input-output dependency given by Eq. (\ref{eq:LIequation}) describes the characteristic s-shaped curve on a double-log scale.
Below threshold, i.e. $n^{\textrm{L}}\ll 1$, spontaneous emission dominates and Eq.~(\ref{eq:LIequation}) becomes linear in $n^{\textrm{L}}$:
\begin{equation}
	P^{\textrm{P}}_{\textrm{abs}}(\lambda^P) \approx \frac{h\nu^ {\textrm{L}}\gamma}{\beta\delta} \biggl[ 1 + (1 - \beta)\xi  \biggr]n^{\textrm{L}}.
	\label{eq:LIequationBelowThreshold}
\end{equation}
\noindent This leads to $P^{\textrm{P}}_{\textrm{abs}}(\lambda^P) \propto (n_{0} / \delta) n^{\textrm{L}}$ for the case of $\beta \ll 1$.
Above threshold, i.e. $1 \ll n^{\textrm{L}}$, similar arguments result in $P^{\textrm{P}}_{\textrm{abs}}(\lambda^P) \propto (1 / \delta) n^{\textrm{L}}$, however, here not relying on any assumption regarding $\beta$.
The input-output curve's slope below and above threshold therefore depend on $\delta$ while $n_0$ modifies the ratio between both, and $\beta$ is sensitive on the nonlinearity at the lasing transition. The different operating regimes of Eq. (\ref{eq:LIequation}), i.e. below, around and above threshold therefore allow determining $n_0$, $\beta$ and $\delta$, given that $\kappa(\lambda^{\textrm{P}})$, $\gamma$ and $\tau_{\textrm{sp}}$ are known.

Here, we use $\tau_{\textrm{sp}}=1~$ns and $Q=1.5\cdot10^4$ (reported in reference \cite{Heuser2020} for identical QDMLs) determines $\gamma = (2\pi \nu^{\textrm{L}})/Q$.
The pump absorption coefficient $\kappa(\lambda^{\textrm{P}})$ was obtained using material absorption, $R^{\textrm{DBR}}(\lambda)$ and QDML dimensions, see Sec. \ref{sec:PumpAbs}.
$\beta$ depends on the QDML cavity-geometry and the spectral overlap between QD-emission and the lasing mode \cite{Gerard1998}, $\delta$ on the exciton relaxation channels.
Our QDMLA is spectrally highly uniform \cite{Heuser2020}, hence with similar QD-emission and lasing wavelength conditions.
A similar local temperature, assuming local heating induced by $P^{\textrm{P}}$ is comparable for the same pumping conditions, should therefore results in comparable $\beta$s for all QDMLs then pumping with the same $\lambda^{\textrm{P}}$.
Exciton relaxation is a property of gain material and pump process, hence $\delta$ should mostly depend on the transition driven by the pump laser.
When fitting Eq. (\ref{eq:LIequation}) to our output curves measured at varying conditions we use the dispersion of fit parameters as validation of our hypotheses.
We therefore study the impact of having $n_0$, $\beta$ and $\delta$ as free fit parameter of each laser, wavelength or only pumping concept.
We pay particular attention to unexpected wavelength dependencies of fit parameters, as this would indicate a shortcoming of our closed-form fitting routine.

\section{\label{sec:Experiment}Experimental setup}

\begin{figure*}[t]
	\centering
	\includegraphics[width=11cm]{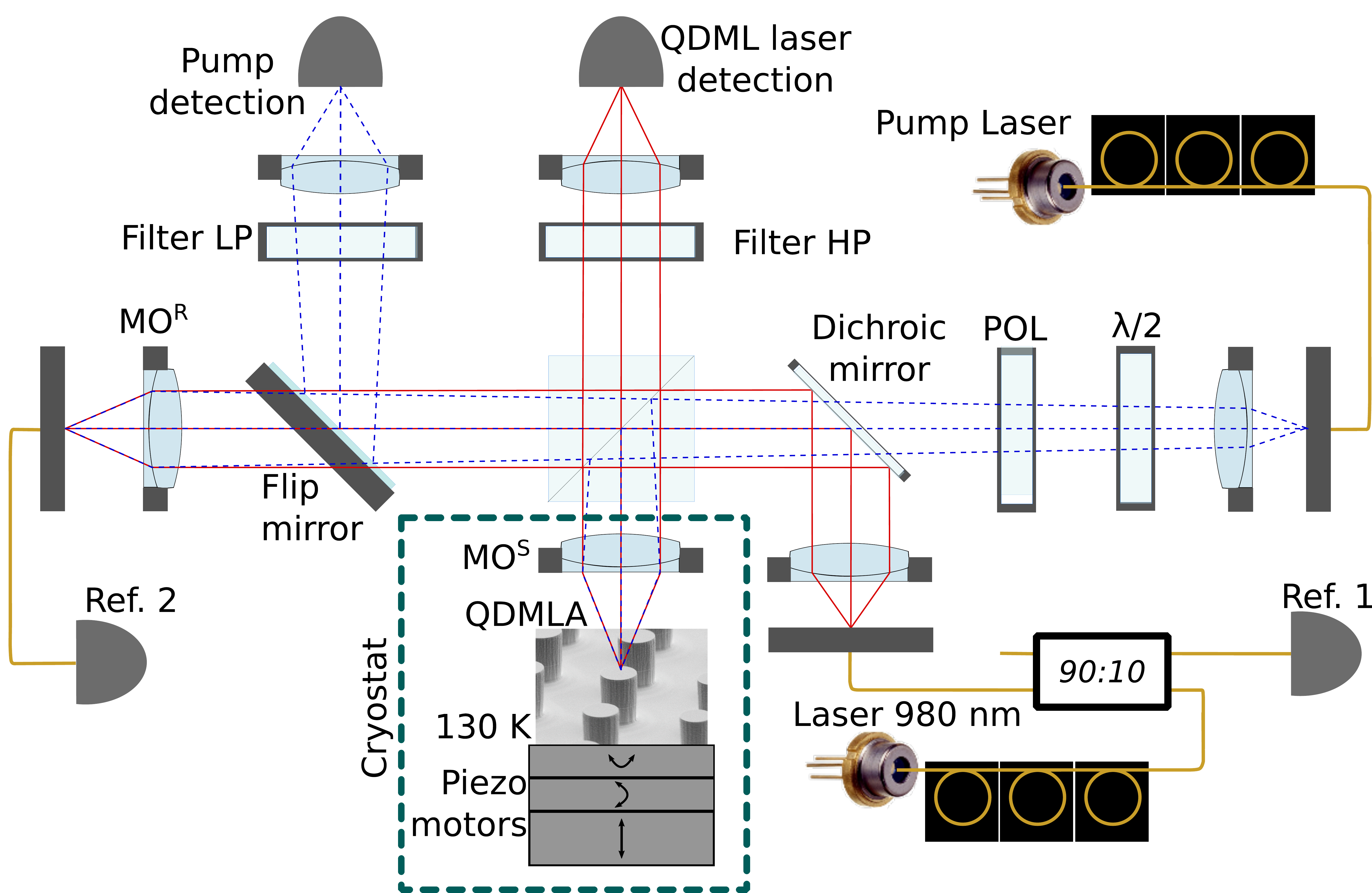}
	\caption{
		Experimental scheme.  Multiple optical references ensure comparable conditions for the wide range of pump wavelengths. A reference microscope objective (MO$^{\textrm{R}}$) and laser (980~nm) reproduce the optical conditions of imaging (MO$^{\textrm{S}}$) of the QDMLA sample.}
	\label{fig:opticalScheme}
	\hrule
\end{figure*}

The QDMLs are experimentally characterized by measuring input-output curves. Here, we record the QDML's power in both polarization states of the bimodal micropillar cavity \cite{Schlottmann2018}.
In the present work, we did not observe polarization switching inside the relevant pump power ranges.
As  illustrated in  Fig.~\ref{fig:opticalScheme}, a half wave-plate ($\lambda /2$) and a polarizer (POL) accurately control the optical pump power $P^{\textrm{P}}$.
The pump laser is detected by a powermeter placed after a short-pass filter (Thorlabs FESH0900) which suppresses the QDMLs intensity by over 50 dB.
The QDML emission is measured in a similar way after three consecutive long-pass filters (Thorlabs FELH0950), which remove crosstalk from the pump-laser by 15 orders of magnitude.
The recordings of both detectors are post-processed to obtain the pump and emission intensity at the top of each QDML.
For that, we take the spectrally dependent transmission coefficients of each component inside the respective beam path into account.

We characterize a total of eight QDMLs located in different regions of the laser array.
The sample is placed inside a liquid nitrogen cryostat, under secondary vacuum at 130 K (measured below the sample).
The sample is mounted on a stack of three piezo-motors for precise alignment of its parallelism and distance relative to the sample microscope objective  ($\mathrm{MO}^{\mathrm{S}}$) focal plane.
Here, we use a MAG10, NA=0.3 objective (Olympus LMPLN10XIR) which we place inside the  cryostat's cold-chamber.
This eliminates optical aberrations otherwise induced when the cryostat's window is located between $\mathrm{MO}^{\mathrm{S}}$ and the sample.
We target the extraction of quantitative values, and we therefore have to pay great attention to the calibration of all involved components and to the creation of reference systems.
A particular challenge is posed by the fact that we vary the pump wavelength by as much as $250~\textrm{nm}$, and any chromatic aberrations have to be compensated.
For this reason we have implemented strict and reproducible alignment criteria detailed below which results in equivalent experimental conditions for the different pump wavelengths.

\begin{table}
	\centering
	\begin{tabular}{| c p{0.1 cm}  || c p{0.5 cm}   c p{0.5 cm}   c p{0.5 cm}   c |}
		\hline
		Wavelength (nm) && 912 && 785 & & 730 && 660 \\
		\hline
		Injection eff. Ref. 2 && - && 0.59 & & 0.57 && 0.64 \\
		\hline
	\end{tabular}
	\caption{Summarize of the MO transmission and the injection coefficients through the reference single mode fiber depending on the wavelength.}
	\label{table:InjectionCoefs}
\end{table}

First, we use a reference laser at 980 nm, the wavelength of the QDMLs' lasing mode.
The laser is coupled to a 90:10 four-port single mode (sm) fiber-splitter.
The 90\% output port is connected to the experiment via collimation optics, which we align using a mirror placed inside the optical path directly following the collimation section.
Mirror angle, the distance between collimation lens (Throlabs AC254-030-B-ML) and single-mode (sm) fiber output, as well as the fiber's $(x,y)$-position are optimized such the collection (measured at the splitter's 10\% output-port) of the mirror's back-reflection is maximized.
Our 980 nm alignment reference (Ref. 1) is therefore collimated and optimized using independent and sensitive confocal detection.
Second, outside the cryostat we emulate optical injection into QDMLs using a sm-optical fiber (Thorlabs 780HP).
To create comparable conditions as for puming QDMLs, the microscope objective (MO$^{\textrm{R}}$) focusing into the reference fiber is identical to the one used for optically pumping the QDMLA (MO$^{\textrm{S}}$).
This section of the setup serves as our second reference (Ref. 2), and it is aligned  by maximizing the collection efficiency of Ref. 1, hence to conditions matching the QDMLs.

The third step is adjusting the sample's position inside the cryostat.
We again use Ref. 1 and ensure that its focal spot is formed on top of each QDML.
For that, the distance between the QDML and MO$^{\textrm{S}}$ was optimized for maximum collection of the 980~nm reference laser reflected off a QDML by the 10\% output port of Ref. 1.
This again leverages confocal detection at the QDMLA's emission wavelength for high absolute alignment accuracy.
As the final step, we align each of the different pump lasers by maximizing their power injected into Ref. 2.
We compensate for chromatic focal distance of $\mathrm{MO}^{\mathrm{S}}$ by adjusting the pump lasers' collimation optics such that collection by Ref. 2 is optimized.
This ensures that each pump's focal spot is located on the top of the QDMLs, despite of their large difference in wavelength.
Only after this careful procedure we can be certain that the pump laser will always be focused in the same plane.

\begin{figure*}[t]
	\centering
	\includegraphics[width=11cm]{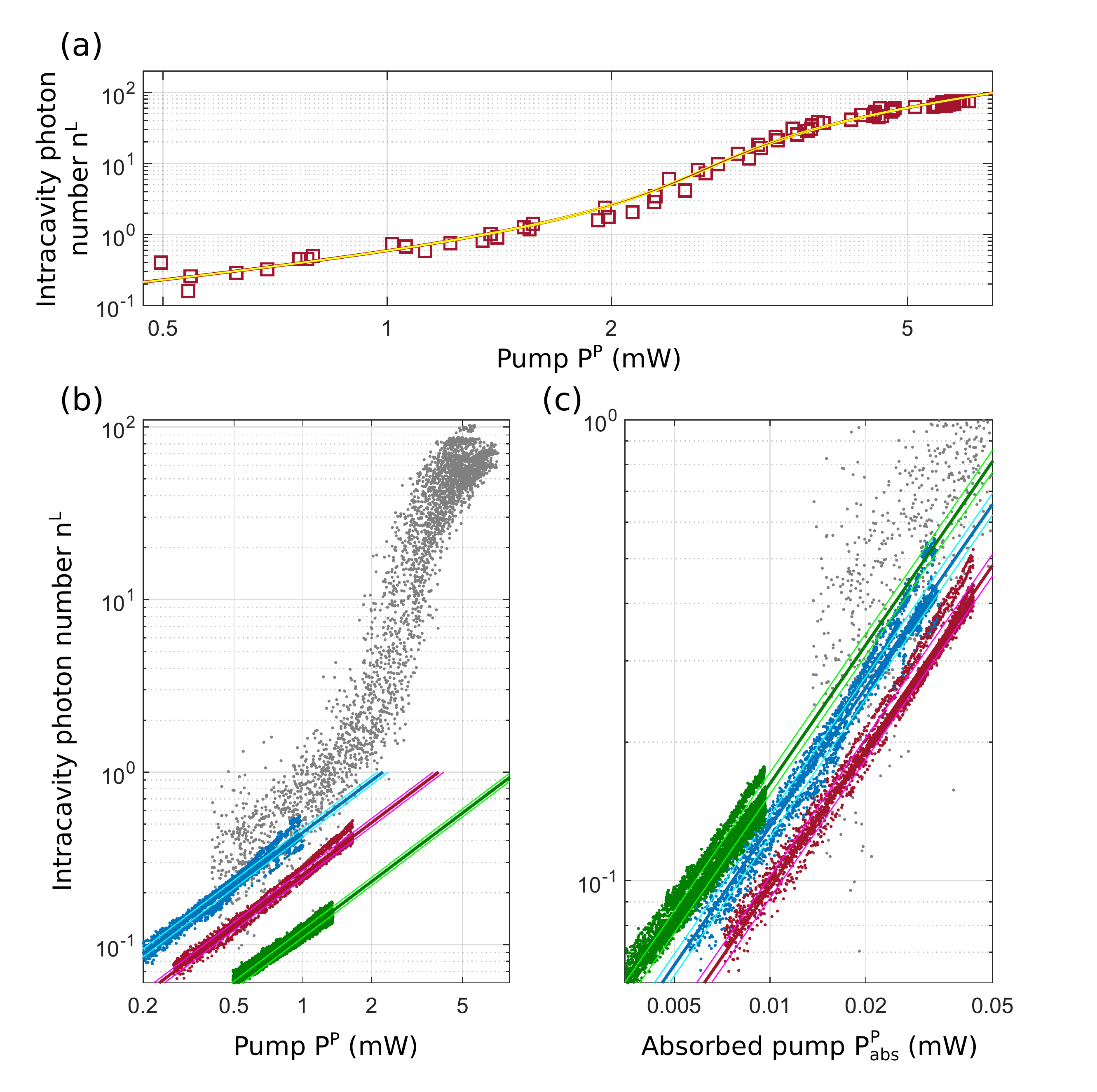}
	\caption{
		(a) Pumping the bulk GaAs of QDMLs results in a lasing threshold of around 1.5~mW, $\beta^{\textrm{GaAs}}=2.2\cdot^{-2}$. (b) Pumping with 786~nm creates text-book input-output curves, while the thermal roll over prohibits lasing when pumping with 730~nm and 660~nm.
		(c) Rescaling to the absorbed pump power accounts for the differences obtained. Data for pumping with 786~nm, 730~nm and 660~nm is given in blue, red and green, respectively.}
	\label{fig:GaAsPump}
	\hrule
\end{figure*}

For confirmation we determine the collection efficiency of Ref. 2 for each pump laser, see table~\ref{table:InjectionCoefs}.
Noteworthy, we obtain highly reproducible collection efficiencies.
This ensures that the alignment between the QDMLA and the different pump lasers is close to identical.

The reflection spectrum of each QDML, average shown in Fig. \ref{fig:DBRSpectrum}, was characterized using a supercontinuum from 650~nm to 1000~nm.
In order to probe the QDMLs under conditions identical to their optical pumping, the supercontinuum is beforehand injected into a 50:50 four-port sm-optical fiber-splitter (Thorlabs TW1064R5A2A).
It is subsequently collimated and focused on the QDMLs, and the reflection is collected by the same fiber-splitter and its spectrum recorded by an optical spectrum analyzer (Yokogawa AQ6370D).
We then inserted a broadband mirror (Thorlabs BB1-03) before $\mathrm{MO}^{\mathrm{S}}$, and used the reflected spectrum to normalize the QDMLs' reflection spectrum.
This removed artifacts originating from chromatic attenuations inside the measurement path.

\section{\label{sec:Results}Results}

\subsection{\label{sec:ResultsBulk}  Bulk GaAs pumping}

By selecting the wide range 660~nm$\leq\lambda^{\textrm{P}}\leq$786~nm to pump the GaAs matrix we test the validity of our approach for a variety of absorption coefficients.
We find that $\alpha^{\textrm{GaAS}}(\lambda^{\textrm{P}})$ has a major influence on the QDMLs' performance.
When pumping at $\lambda^{\textrm{P}}=786~$nm, we obtain an average lasing threshold of $P^{\textrm{P}} = (1.5 \pm 0.12)$~mW ($\approx21~\textrm{kW/cm}^2$), see Fig. \ref{fig:GaAsPump}(a) for an example input-output curve with an exemplary fit to Eq. (\ref{eq:LIequation}) using $\kappa^{\textrm{GaAs}}(786~\textrm{nm})=1$.
However, for shorter pump wavelength absorption by the top DBR increases, resulting in non-radiative decay that locally heats the QDMLs, reducing the internal quantum efficiency and causing thermal rollover of the input output characteristics \cite{Agrawal1993, Baveja2011}.
For $\lambda^{\textrm{p}}=730~$nm and 660~nm this effect is so strong that it prohibits the QDMLs to cross the lasing threshold.
A comparison between the three different GaAs pump wavelenghts is given in Fig. \ref{fig:GaAsPump}(b).
The gray dots show the collective data of 40 input-output curves obtained for $\lambda^{\textrm{P}}=786~$nm.
The blue, red and green data correspond to below threshold emission for $\lambda^{\textrm{P}}=786~$nm, $\lambda^{\textrm{P}}=730~$nm and $\lambda^{\textrm{P}}=660~$nm, respectively.
Here, we used a more sensitive photodetector in order to obtain a higher signal to noise ratio at the nW power-levels below threshold.

The analysis of the experimental data starts with fitting Eq. (\ref{eq:LIequation}) to the data obtained for $\lambda^{\textrm{P}}=786~$nm using $\kappa^{\textrm{GaAs}}(786~\textrm{nm})=3.34\cdot10^{-2}$.
Allowing for an independent $n_0$ per QDML and one global $\beta$ as well as $\delta$, we obtain  $\beta^{\textrm{GaAs}}(786~\textrm{nm})=(2.24 \pm 0.03)\cdot10^{-2}$, $\delta^{\textrm{GaAs}}(786~\textrm{nm})=(1.9 \pm 0.01)\cdot10^{-2}$ and $n_{0}^{\textrm{GaAs}}(786~\textrm{nm})=2913 \pm 70$.
$\beta$ and $\delta$ are determined with an uncertainty in the \% range, and $n_0$ has a comparable standard deviation across the eight lasers.
The $\beta$ factor is within the range commonly determined in other experiments, and we find a very low exciton conversion efficiency $\delta$. Only a couple of percent of excitons created by the pump therefore 
end up in a QD transition contributing to lasing.

For $\lambda^{\textrm{p}}\leq 730~$nm our QDMLs do not lase, yet data recorded below threshold allows us to confirm our findings.
Based on the calculated $\kappa^{\textrm{GaAs}}(\lambda^{\textrm{P}})$ we re-scale the input-output curves for all GaAs pump wavelengths and show the resulting data in Fig. \ref{fig:GaAsPump}(c).
In accordance to the arguments outlined in Sec. \ref{sec:FreePara}, we now define $n_0$ as a laser parameter independent of the GaAs pump wavelength.
We keep $\beta=2.24\cdot10^{-2}$ constant at the value obtained for $\lambda^{\textrm{P}}=786~$nm, and for this small value it's impact is mostly negligible below threshold.
Finally, we set $\delta$ to be independent of the QDML, yet a free parameter for all three pump wavelengths.
From the linear fitting of the curves depicted in Fig. \ref{fig:GaAsPump}(c) via Eq. (\ref{eq:LIequationBelowThreshold}), we obtain $n_{0}^{\textrm{GaAs}}=3496 \pm 254$ and $\delta^{\textrm{GaAs}}(786~\textrm{nm})=2.4\cdot10^{-2}$, $\delta^{\textrm{GaAs}}(730~\textrm{nm})=1.8\cdot10^{-2}$ and $\delta^{\textrm{GaAs}}(660~\textrm{nm})=3\cdot10^{-2}$.
These parameters and Eq. (\ref{eq:LIequationBelowThreshold}) approximate the QDMLs' sub-threshold dependency very well, as is apparent from data (dots) and fits (lines) in Fig. \ref{fig:GaAsPump}(c).
Crucially, the very small variations in $\delta^{\textrm{GaAs}}$ mean that the calculated $\kappa(\lambda^{\textrm{P}})$ accounts for the QDMPA $\lambda^{\textrm{P}}$ dependence when pumping GaAs.
Also, we expected essentially identical $\delta^{\textrm{GaAs}}$s as the three pumping levels are linked through efficient intra-band relaxation.

\subsection{\label{sec:ResultsWL} Wetting layer pumping}

\begin{figure*}[t]
	\centering
	\includegraphics[width=11cm]{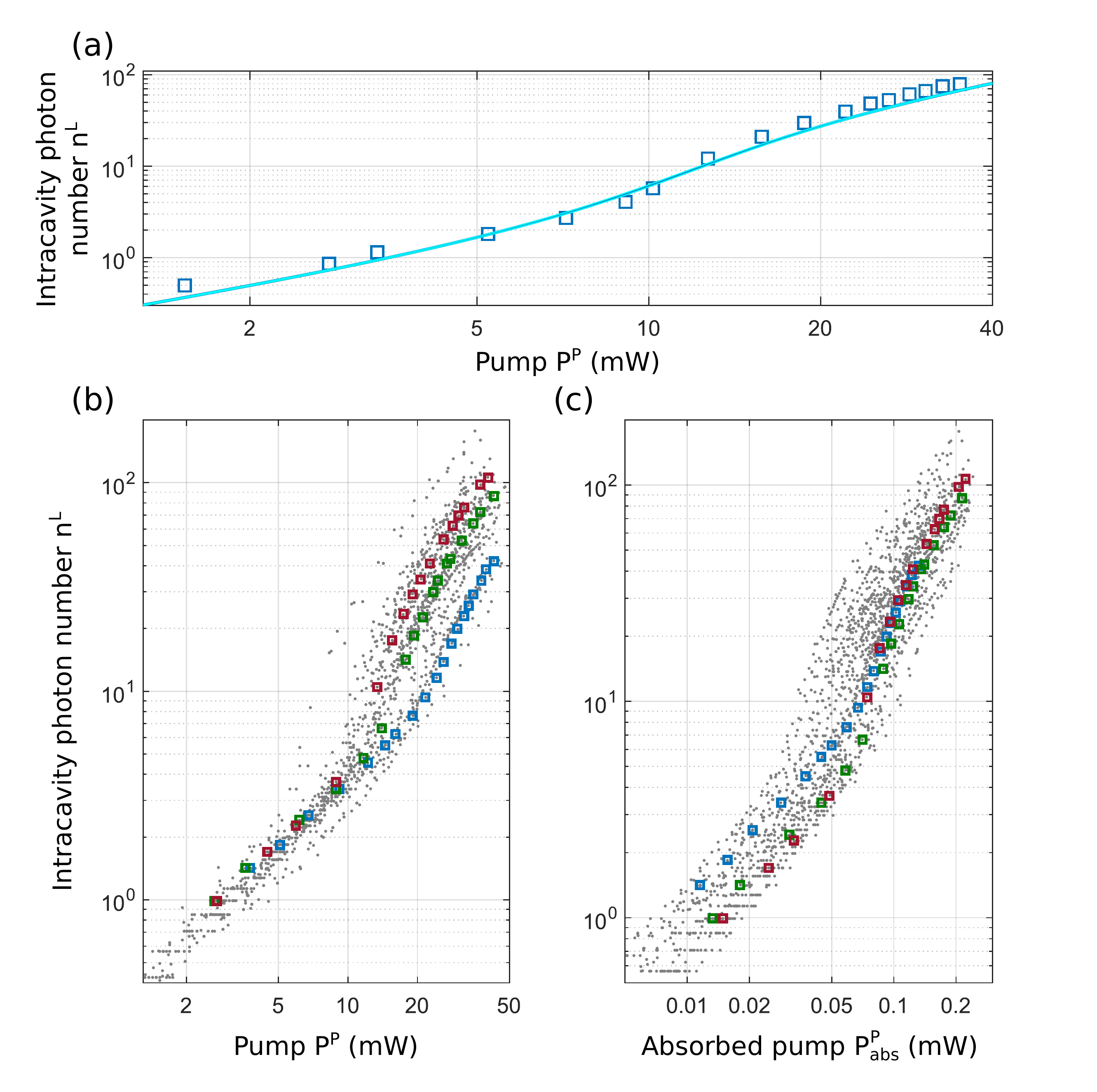}
	\caption{
		(a) Input output curve for pumping the WL at maximum DBR transmission,$\beta^{\textrm{WL}}=9.7\cdot^{-2}$.
		(b) We recorded >100 data sets for eight different lasers with approximately 6 different $\lambda^{\textrm{P}}$ for each QDML. The square data shows corresponds to the same pillar with $\kappa=3.1\cdot10^{-3}$ (blue), $\kappa=5\cdot10^{-3}$ (green) and $\kappa=5.5\cdot10^{-3}$ (red). (c) Plotting the same data using $P^{\textrm{P}}_{\textrm{abs}}$ compensates for the strong variations.}
	\label{fig:WLPumping}
	\hrule
\end{figure*}

We characterized pumping in the range from 905 nm to 922 nm, which spans the DBR reflection minimum indicated by the dashed line in Fig. \ref{fig:DBRSpectrum}.
Crucially, for a sample temperature of $\sim130~$K we expect the WL's transition within this range.
A typical intput-output curve obtained at the DBR's transmission maximum at $\lambda^{\textrm{P}}=912.3~$nm is given in Fig. \ref{fig:WLPumping}(a).
We recorded a total of 114 input-output curves for the eight QDML, and initially we treated $\kappa^{\textrm{WL}}$ as constant across the entire WL pump range.
According to our previous line of argument, we defined a single $\beta^{\textrm{WL}}$ and $\delta^{\textrm{WL}}$ for fitting all data, but allowed for one $n_0$ for each input-output curve, hence of each laser-$\lambda^{\textrm{P}}$ pair.
Post-treatment of $n_0^{\textrm{WL}}(\lambda^{\textrm{P}})$ by sliding window averaging (width 12) revealed a strong dependency on $\lambda^{\textrm{P}}$.
The spectral profile closely resembles a convex polynomial centered at $\lambda^{\textrm{P}}=915~$nm and with a full width at half maximum of 14.8 meV.
This is comparable to the $\sim10~$meV typical for the WL\cite{Ksendzov1991}, and we link this spectral feature to the WL's absorption.
According to Eq. (\ref{eq:LIequationBelowThreshold}), both $n_0$ and $\kappa$ are linearly related, and hence we include the WL's absorption spectrum by multiplying $\kappa^{\textrm{WL}}(\lambda^{\textrm{P}})$ with the normalized $n_0^{\textrm{WL}}(\lambda^{\textrm{P}})$ dependency.

For one QDML we illustrate the impact of the WL's absorption spectrum.
The red, green and blue squares in Fig. \ref{fig:WLPumping}(b) belong to the same QDML pumped at values with different $T^{\textrm{DBR}}$, and the three input-output curves strongly differ in amplitude when viewed as a function of pump power $P^{\textrm{P}}$.
However, all three excellently agree when we change the abscissa to $P^{\textrm{P}}_{\textrm{abs}}$ calculated with the WL's absorption profile included, see Fig. \ref{fig:WLPumping}(c).
We again fit all 114 input-output curves globally with the new $\kappa^{\textrm{WL}}(\lambda^{\textrm{P}})$ using one common $\beta^{\textrm{WL}}$ and $\delta^{\textrm{WL}}$, yet with an independent $n_{0}$ for each data.
Again, the obtained $\beta^{\textrm{WL}}=(9.7 \pm 0.2)\cdot10^{-2}$, $\delta^{\textrm{WL}}=(7.5 \pm 0.07)\cdot10^{-3}$ and $n_{0}^{\textrm{WL}}=627 \pm 151$ are very concise and confirm the validity of our approach.

\subsection{\label{sec:Absorption} Effective pump absorption}

\begin{figure*}[t]
	\centering
	\includegraphics[width=11cm]{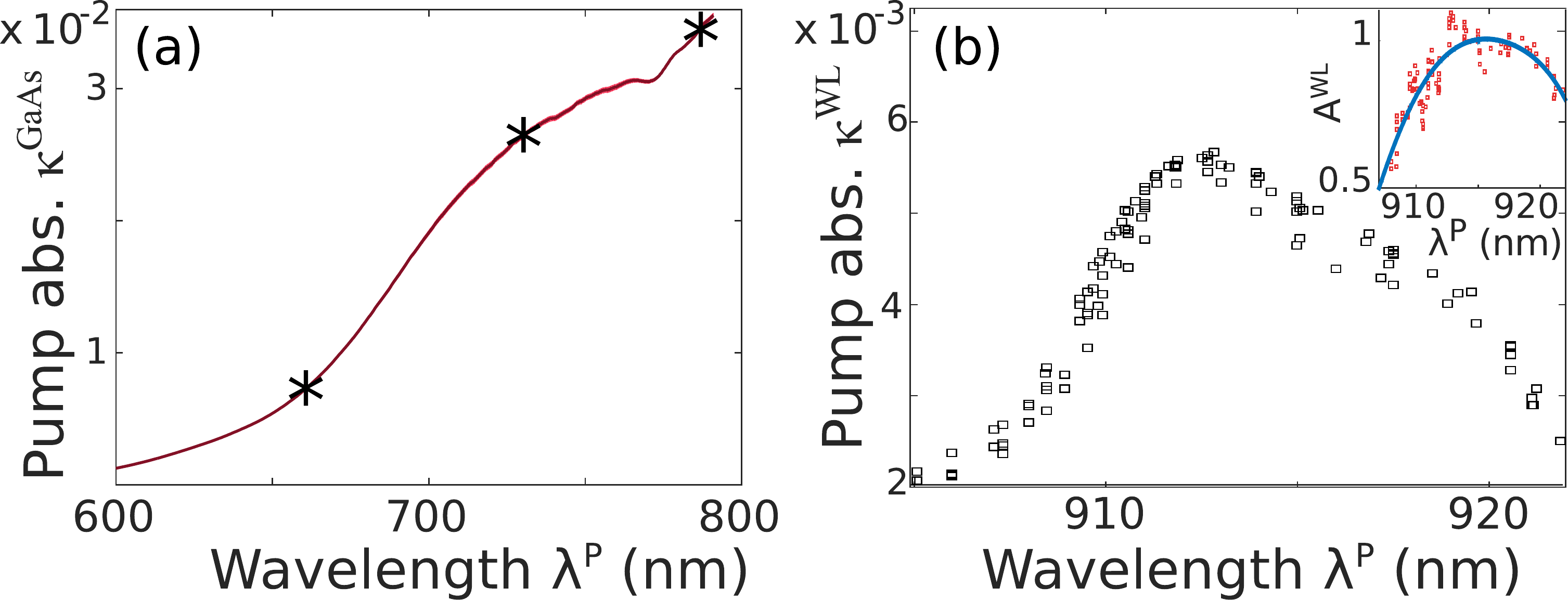}
	\caption{
		(a) Pump absorption $\kappa^{\textrm{GaAs}}$ by the bulk GaAs transition, obtained from $R^{\textrm{DBR}}(\lambda)$ and $\alpha^{\textrm{GaAs}}(\lambda)$. 
		(b) Pump absorption $\kappa^{\textrm{WL}}$ for WL pumping. Inset shows the WL's extracted absorption spectra.}
	\label{fig:AbsSpectra}
	\hrule
\end{figure*}

During the previous sections we were able to determine and confirm the effective pump absorption $\kappa(\lambda^{\textrm{P}})$ as an essential parameter.
In the case of bulk GaAs pumping the combination of $\alpha^{\textrm{GaAs}}$ and $R^{\textrm{DBR}}$ resulted in $\kappa^{\textrm{GaAs}}(\lambda^{\textrm{P}})$ shown in Fig. \ref{fig:AbsSpectra}(a).
The three stars correspond to the three GaAs pumping conditions with $\lambda^{\textrm{P}}=660~$nm, $\lambda^{\textrm{P}}=730~$nm and $\lambda^{\textrm{P}}=785~$nm.
Figure \ref{fig:AbsSpectra}(b) shows $\kappa^{\textrm{WL}}(\lambda^{\textrm{P}})$ for the eight QDML, and the inset gives the indirectly determined and normalized WL absorption spectra.

At some stage we attempted to increase the absorption efficiency $\kappa$ by focusing the pump laser waist at the top of the GaAs cavity.
The idea was to minimize the impact of DBR absorption by partially pumping the QDMLs through their side walls.
Regarding GaAs pumping, the associated results showed no significant improvement for $\lambda^{\textrm{P}}=786~$nm, and even a slight degradation for $\lambda^{\textrm{P}}=730~$nm and $\lambda^{\textrm{P}}=660~$nm.
However, the margin of these modifications were too small to arrive at a conclusion.
Ultimately, the highest pumping efficiency of the WLs occurs when the pump beam waist is focused on the top of the QDMLs.

\section{\label{sec:Conclusion}Conclusion}

We have characterized with great detail the input-output curves of eight QDMLs from an array with homogenized emission wavelengths.
Considering the free parameters one can determine using a simple rate-equation description of QDMLs, we measured the QDML's spontaneous emission confinement factor $\beta$, the exciton conversion efficiency $\delta$ and the number of excitons at transparency $n_0$.
We lend significant robustness to our results by fitting a large number of QDMLs at a variety of pumping conditions.
Parameter $n_0$ was individually optimized for each data or laser, while $\beta$ and $\delta$ were kept uniform for a particular laser or pumping condition.
For the correct global description of optical pumping, $n_0$ showed no dependency of the particular pumping wavelength, and we achieved this for describing GaAs and WL pumping alike.

Our study focuses on quantifying the fundamental mechanisms of QDML optical pumping and through that to clearly identify potential future improvements.
We place one focus on the top DBR's absorption and link it to its intrinsic material composition, the sample temperature and pump wavelength.
These are either well documented or readily measurable quantities, and we find that absorption by the top DBR substantially lowers pumping efficiency as excitons created inside its layers are prohibited from tunneling to the lasers' active region.
Noteworthy, these limitations do not only apply to micropillars or lasers based on QD gain material, but are generally relevant for optical pumping in vertical-cavity surface-emitting lasers (VCSELs) inthe GaAs material system. 
Furthermore, we find that the exciton conversion efficiency $\delta$ of pumping the bulk GaAs is significantly higher than when pumping the WL.
Initially this is surprising: the WL is in direct contact with the QDs, and intuitively we expected the opposite ratio.
However, excitons in the GaAs matrix can relax into the WL and the QDs, and this combined recombination channels could be the reason for $\delta^{\textrm{GaAs}} > \delta^{\textrm{WL}}$.
Finally, we identified a low exciton conversion efficiency in general, which indicates significant contributions from non-radiative decay.

WL-pumping is fundamentally limited to below 1\% power efficiency.
While currently the efficiency when pumping bulk GaAs is limited to a similar range, its is not consequence of a fundamental limitation.
The responsible absorption by the top DBR can be mitigated by DBR-layers comprising of materials with a bandgap higher than the one of the QDML's cavity, for example substituting the DBR's $\textrm{GaAs}$ layers for $\textrm{Al}_{0.1}\textrm{Ga}_{0.9}\textrm{As}$.
This would elevate their bandgap by 125~meV, while the modified refractive index increase would reduce the reflectively by less than $0.5$\% only for the same number of mirror pairs.
This should increase pump absorption in the laser gain section by a factor of $7.4$, thus exceeding $20$\%.
Even higher Al content would enable shorter pump-wavelengths, which in turn would further increases the pump absorption in the central GaAs cavity.
That only a couple \% of excitons created by the pump inside the gain medium do contribute to the lasing process might be due to several reasons, and a contribution by non-radiative decay at defects located at the QDML's etched sidewalls could be mitigated through surface passivation \cite{Higuera-Rodriguez2017}.
$\textrm{GaAs}$ passivation by nitridation has demonstrated to reduce the surface state density by a factor of 6, and the effect appears stable over time \cite{alekseev2015nitride}.
Such improvements using established techniques therefore  promise a significant improvement of QDML's conversion efficiency by more than one order of magnitude.

Our results are relevant for the growing field of dense laser arrays.
Vertically emitting laser arrays are relevant for innovative applications such 3D sensors \cite{Seurin2016}, structured light, optical free-space communication \cite{liverman2019vcsel} and optical neural networks \cite{Brunner2013a,Brunner2015,Heuser2020}.
Individual electrical contacts for each laser needs to allocate a large fraction of a chip's surface area to electrical wires, and as a consequence the number of devices in such laser arrays remain small, usually limited to <100 lasers \cite{Brunner2015,Heuser2020}.
Free-space pumping techniques do not suffer from this limitation and hence can alleviate this bottleneck in the general context of research \cite{Heuser2018} but also in applications \cite{guina2017optically,sills2013phase}.
Furthermore, high speed electrical modulation is prohibited due to the capacity of long electrical connections in large arrays.
Fast modulation through the optical pump allows can address this challenge by shifting the modulating infrastructure outside the laser array.
In all these aspects pump efficiency is of major importance, as the currently low efficiency quickly limits to number of lasers for the practical range of 10s of mW pump power.
Finally, while our work uses free-space techniques, it is directly transferable to optically pumped QDMPAs integrated in a waveguide circuitry.
In such a setting, adding additional waveguides for the optical pump would not require additional fabrication steps, which could give it an additional advantage over electrical pumping.

\section*{Funding}
The authors acknowledge the support of the Region Bourgogne Franche-Comt\'{e}.
This work was supported by the EUR EIPHI program (Contract No. ANR-17-EURE- 0002) and by the Volkswagen Foundation (NeuroQNet I\&II).
X.P. has received funding from the European Union’s Horizon 2020 research and innovation program under the Marie Skłodowska-Curie grant agreement No 713694 (MULTIPLY).

\section*{Disclosures}
The authors declare no conflicts of interest.

\bibliography{bibliography}

\end{document}